\documentclass{article}
\usepackage[utf8]{inputenc}
\usepackage[a4paper,margin=1.25in,headheight=15pt]{geometry}

\usepackage[font=footnotesize]{caption}
\usepackage{subcaption}
\usepackage{graphicx}
\usepackage{setspace}
\usepackage{xcolor}
\usepackage{hyperref}

\graphicspath{ {figures/} }

\usepackage{amsmath}
\usepackage{amsfonts}
\usepackage{amssymb}
\usepackage{amsthm}
\usepackage{mathtools}

\usepackage[linesnumbered,ruled]{algorithm2e}
\SetKwIF{If}{ElseIf}{Else}{if}{}{else if}{else}{end if}%
\SetKwFor{While}{while}{}{end while}%
\SetKwRepeat{Do}{do}{while}
\SetKw{KwGoTo}{go to}

\title{Modeling Human Dynamics and Lifestyle using Digital Traces}
\author{Sharon Xu$^1$ \and Steven Morse$^2$ \and Marta C. Gonz{\'a}lez$^3$}
\date{%
    $^1$Massachusetts Institute of Technology, Cambridge, MA \\
    $^2$United States Military Academy, West Point, NY \\
    $^3$University of California, Berkeley, CA\\[2ex]
    March 2020
}

\begin{document}

\maketitle

\begin{abstract}
Human behavior drives a range of complex social, urban, and economic systems, yet understanding its structure and dynamics at the individual level remains an open question.  From credit card transactions to communications data, human behavior appears to exhibit bursts of activity driven by task prioritization and periodicity, however, current research does not offer generative models capturing these mechanisms.  We propose a multivariate, periodic Hawkes process (MPHP) model that captures --- at the individual level --- the temporal clustering of human activity, the interdependence structure and co-excitation of different activities, and the periodic effects of weekly rhythms.  We also propose a scalable parameter estimation technique for this model using maximum-aposteriori expectation-maximization that additionally provides estimation of latent variables revealing branching structure of an individual's behavior patterns.  We apply the model to a large dataset of credit card transactions, and demonstrate the MPHP outperforms a non-homogeneous Poisson model and LDA in both statistical fit for the distribution of inter-event times and an activity prediction task.
\end{abstract}



\section{Introduction}

Within the last decade, the digital age has sharply redefined the way we study human behavior. Electronic records now encompass a diverse spectrum of human activity, ranging from phone \cite{ccall1,ccall2,stmorseieee} and email communication data \cite{poissonianemail} to location records \cite{cmobility_limits,cmobility} and household energy usage \cite{energykwac,energyxu}. The existence of these passively collected datasets supersedes the need to actively collect data through cumbersome and expensive surveys. With the rising ubiquity of passive data, we now have new opportunities to understand the individual dynamics at a higher level of granularity in time scale and behavioral detail. Models of human behavior have the potential to inform government policy, helping to optimize infrastructure, reduce congestion, and mitigate pollution. 

In particular, the analysis of credit card records can give us a fine-grained understanding of spending patterns, lending valuable insight into the design of cities, the distribution of wealth, and the urban economy. The primary use of this emerging data source has centered around measuring similarity in purchases through affinity algorithms \cite{affinity1, affinity2}. However, recent work has started to link both mobility and socio-demographics with purchase behavior \cite{rdicle, socialbridges}, suggesting that models of the individual have broader implications. For example, although traditionally mobile phone data is the basis for models of human mobility, recent research has shown that credit card data enables similar applications. That is, through the preferred transitions between businesses, we can model movement resulting from shopping activity, observing the same imbalance in the spatial distribution as found in traditional mobility studies \cite{ccmobility}.

Despite the wide range of applications, current literature lacks temporal models of shopping patterns. Indeed, there are few urban models describing human dynamics at the individual level in present research. One exception is a recent framework for urban mobility, \texttt{TimeGeo} \cite{timegeo}. This framework learns a high resolution model from passive data containing sparse traces of individuals. It explicitly outlines the choice mechanisms each individual makes, delineating a procedure in which, for example, individuals start at home, then must choose whether to move, then choose whether to explore, and then choose whether to visit a previously unvisited location.

In this paper, we propose a unified statistical framework to describe multidimensional human dynamics.  Applying our method to credit card transaction histories, we model the individual dynamics of shopping behavior for the first time.  From massive amounts of passive data, the proposed framework extracts the key underlying mechanisms driving human behavior.  The resulting model is simple and interpretable, but comprehensive enough to generate realistic trajectories. As a result, the method lends insight into periodic patterns and temporal transitions in urban areas.  Unlike \texttt{TimeGeo}, we impose no explicit choice mechanisms on behavior, capturing a comprehensive representation of human dynamics through just two well-known behavioral characteristics: burstiness and periodic effects.

A pattern of temporal clustering --- that is, long periods of low activity punctuated by short periods of high activity --- seems to define many natural and human-centric phenomena, from earthquakes and neural impulses to social systems, technological advances, and economic markets.  This so-called ``burstiness'' has been shown to be a fundamental property of human dynamics \cite{poissonianemail,heavytailsbarabasi,hawkes1d,powerlaw}. Past work attributes the burstiness of human dynamics to two driving mechanisms: inherent correlations due to decision-making mechanisms like task-prioritization, and underlying temporal correlations like circadian rhythms (e.g. the home-work-home cycle). 


There is a rich literature surrounding the hypothesis that task prioritization results in the bursty signals in human behavior \cite{barabasi2005origin}. This hypothesis stems from the idea that certain activities with shared prioritization will occur in bursts, resulting in short inter-event times, followed by long periods of inactivity. For example: a taxi ride may result in restaurant and department store transactions; regular payments for network, phone, and cable services may often be made together; a person running weekly errands may make many purchases in a short time period.  Similar patterns occur in communication networks \cite{stmorsethesis}: a call from mother to son may excite a call from son to father; an email from the manager may induce increased communication between team members. Capturing the structure of these excitation patterns gives important insight into individual priorities and behavioral patterns, extending current work, which only considers temporal behavior without the context of different activities.

Queuing process models are a primary example of models based on this task prioritization mechanism. These models attribute burstiness to the execution of tasks based on their priority \cite{prioritylist,barabasi2005origin}, describing the waiting time of a task, or in other words, the time period before a task is executed. This depends on the cumulative time needed to perform all tasks before it. The queuing mechanism and priority distribution are chosen to produce heavy tails in the waiting time distribution \cite{waitingtime,heavytailsbarabasi}, as this has been found to agree well with empirical observations \cite{priorityeinstein}.

A related modeling paradigm to capture the temporal clustering observed in human behavior uses stochastic (point) processes with non-homogeneous rates.  For example, one may engineer the probability of an event occurrence to depend on the occurrence of other, recent events --- this is the so-called ``self-exciting'' property of models like the \emph{Hawkes process} \cite{hawkes1d,powerlaw}. Alternatively, one may modify the process's intensity to fluctuate according to some observed seasonal, weekly, or daily rhythm.  This pattern is evident in activity ranging from location data \cite{cmobility_limits,cmobility}, phone \cite{ccall1,ccall2} and email communication \cite{poissonianemail} to Twitter activity \cite{ctwitter} and open-source contributions on Wikipedia and OpenStreetMap \cite{cwikipedia,copenstreetmap}.  A number of factors contribute to this periodicity, including the day-night cycle, employment status, work schedules and commuting patterns \cite{cemployment, mobilitycities}, and the activity of one’s social contacts \cite{toole2015coupling}.

Perhaps surprisingly, this periodicity appears to provide sufficient explanatory power to model the temporal correlation observed in human behavior, even using memoryless models.  For example, \cite{poissonchangerate} shows that the observed power law scaling on inter-event time distributions can be achieved using Poissonian agents with varying rates. Extending upon this line of work, \cite{poissonianemail} argues that such distinctly non-homogeneous event sequences are solely due to circadian rhythms, proposing a non-homogeneous Poissonian model using ``cascades'' of processes corresponding to the hour and day of week. Extensions of this model uses a Markov process with multiple states to modulate transitions between Poisson models with different rates, thus reflecting periods of high and low activity seen in human communications \cite{markovpoisson,markovpoisson3}.  However, while these methods give a close approximation to observed data, they require a large number of parameters to specify the start and end points for distinct time intervals that represent active periods. Thus the result is not only computationally expensive, but is also not descriptive and gives no generative explanation for diversity in human dynamics \cite{book}. In addition, research \cite{ccall1,notperiodic} has shown that even after removing periodic effects, signals remain bursty.


In this paper, we propose a stochastic model called the multivariate periodic Hawkes process (MPHP) that explicitly captures both the structure of interdependency between multiple types of human activity and the fluctuations in activity rates due to circadian rhythms.  We demonstrate, using a large dataset of credit card transactions, that the MPHP model reveals an inherent branching structure to the observed human behavior which allows interpretation of activity ordering and prioritization.  We show the MPHP provides more predictive power than a periodic Poisson process or a latent Dirichlet allocation model.  We also present a maximum-aposteriori (MAP) expectation-maximization (EM) method for parameter estimation of the MPHP that scales to large datasets and allows use of priors to provide the necessary regularization.

\begin{figure}[tb]
    \centering
    \includegraphics[width=\linewidth]{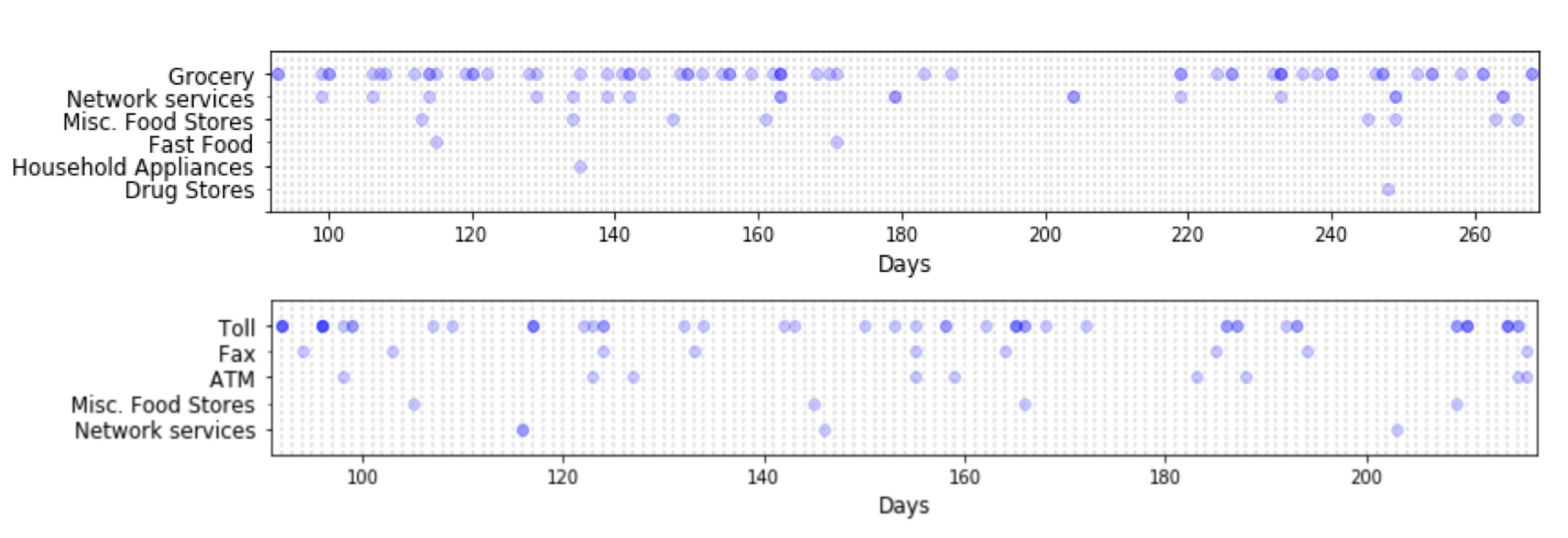}
    \caption{\textbf{Temporal clustering.}  Transaction behavior for two sample users (\textit{top} and \textit{bottom}) exhibits burstiness.  We propose that this pattern can be described by (i) interdependence between activity types and (ii) periodicity and circadian rhythms.}
    \label{userdata}
\end{figure}

\begin{figure}[tb]
    \centering
    \includegraphics[width=0.8\linewidth]{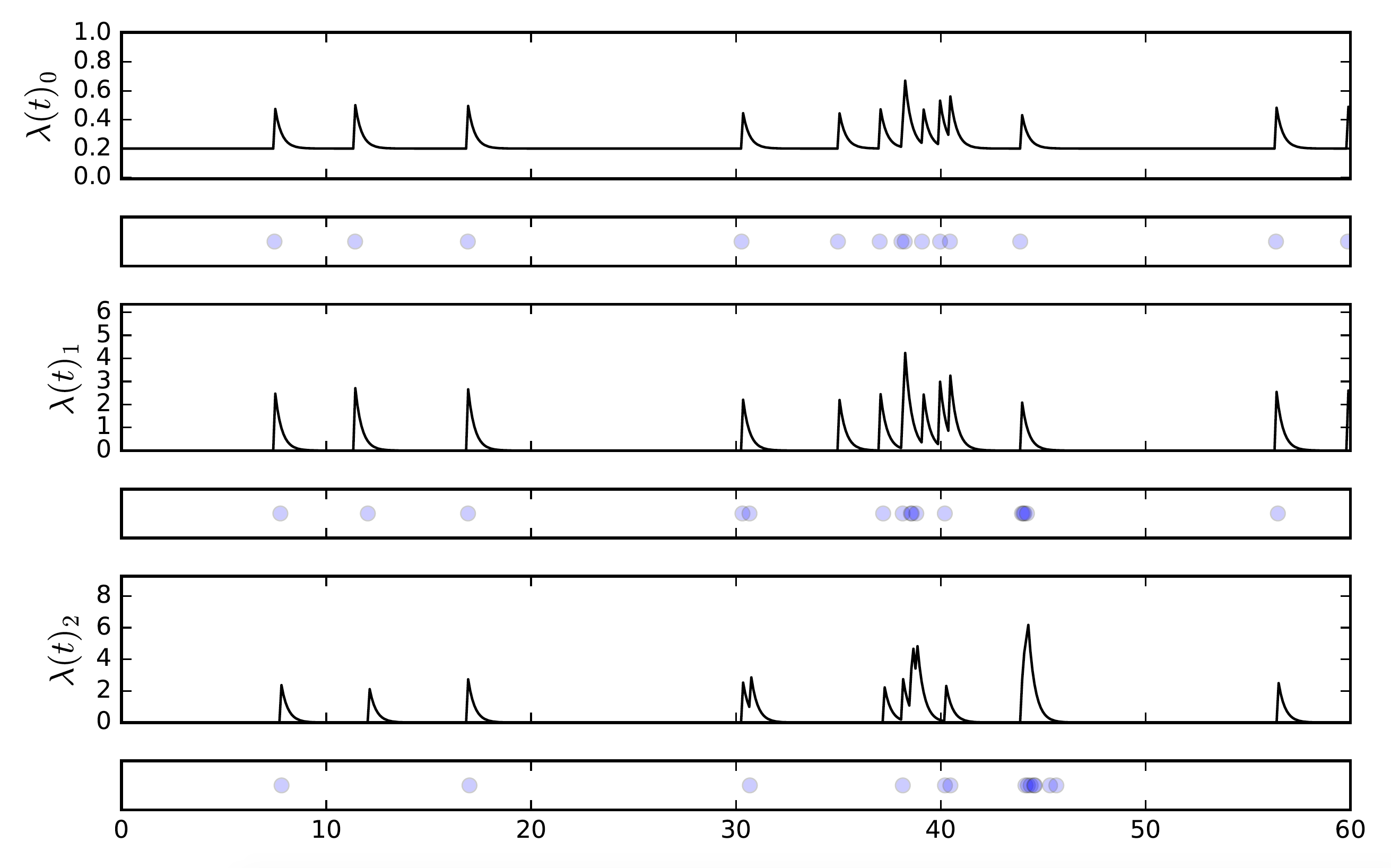}
    \caption{\textbf{Multivariate Hawkes process example.}  Consider the case of $U=3$ activity types, where Activity 0 occurs at some base rate $\mu_0=0.2$, with some self-excitation effect on itself, and also tends to lead to occurrences of Activity 1, which in turn often triggers Activity 2.  This situation is well-modeled by a Hawkes process, with one sample from such a process depicted above.  Notice the cascading effects of Activity 0 to Activities 1 and 2, and temporal clustering due to both self-excitation and co-excitation.}
    \label{example}
\end{figure}


\section{Model}

We propose a multivariate periodic Hawkes process (MPHP) model that captures the rates of different activities, the branching structure of ordering and interdependence of one activity on another, and the periodicity due to weekly circadian rhythms.  We give a brief exposition of the model here, with technical details focused on parameter estimation and simulation to be found in \textit{Methods}.

The Hawkes process \cite{Laub2015} is a point process that is conditionally Poisson, with the conditional rate or intensity $\lambda(t)$ depending on the history of events up to time $t$.  The intensity consists of a background rate and additive contributions from previous events which decay over time.  This allows the model to capture self-excitation and interdependence between events of different types.  We may further incorporate periodicity into the model by scaling the background rate.

Specifically, consider a sequence of events $\{(t_i, u_i)\}_{i=1}^N$ where the $i$th event occurs at time $t_i$ over an interval $[0,T]$ and is of event type $u_i$, out of $U$ total types.  We then define the conditional intensity for each dimension $u$ as
\begin{equation}
    \lambda_u (t) = \mu_u \delta_{d(t)} + 
    \sum_{i \ : \ t_i < t} h_{u_i, u}(t - t_i; \Theta)
\label{eq:mphp}
\end{equation}
where $\mu_u$ is the background intensity for occurrences of type $u$, adjusted by global parameters capturing daily and hourly periodicity.  Each $\delta_d$ scales the background rate depending on the day of the week $d(t)$ corresponding to time $t$.  We let $d=1,...,D$ and ensure these parameters are normalized so that $\sum_d \delta_d = D$.  We use $D=7$, but one may choose more or less granularity to capture, for example, seasonal or hourly effects.

The function $h_{u_i, u}(t_i - t)$ is a \emph{triggering kernel} which represents the excitation effects of an event from $u_i$ at time $t_i$ on the intensity in $u$.  We decompose the triggering kernel into an excitation parameter and decay function:
\begin{equation}
    h_{u_i, u}(t - t_i) = \alpha_{u_i, u} \omega e^{-\omega (t - t_i)}.
\label{eq:trigger}
\end{equation}
We use the common choice of an exponential decay, and set $\omega$ as a single global parameter governing the speed of decay for an individual.  The excitation parameters $\alpha_{ij}$ capture the level of effect an event in $i$ has on the probability of an event occurrence in $j$, and we may consider the entire excitation matrix $A=[\alpha_{ij}]$.  

As an example of the Hawkes process with a small ($U=3$) number of activities, consider Fig. \ref{example}.  Here Activity 0 has intensity $\lambda_0$ with base rate $\mu_0=0.2$, while Activities 1 and 2 have base rate 0.  We set $\alpha_{0,0} > 0$, $\alpha_{0,1} > 0$, $\alpha_{1,2}$, and all other $\alpha_{\cdot,\cdot}=0$, so that Activity 0 has some self-excitation effect, and there is a cascading effect from occurrences of Activity 0 to occurrences of Activity 1 and then 2.  The figure depicts one sample from such a process.


In order to frame both our results and our parameter estimation methodology, it is also important to understand an interpretation of the Hawkes process as a \emph{branching process} \cite{Laub2015}.  Note that when the intensity $\lambda_u(t) = \mu_u$, we can consider any arrivals as parent events.  But any immediately subsequent event (where now $\lambda_u(t) > \mu_u$ due to the contribution of $h_{u_i,u}(\cdot)$) is either another parent event, or (more likely) an offspring that was a result of a previous parent event’s increase in the intensity function.  Under this interpretation, $\alpha_{u_i,u} > 0$ controls the branching ratio, or likelihood of an arrival causing another arrival.  (Indeed, in order for the process to be stationary we must ensure the largest eigenvalue of the excitation matrix $A$ is $<1$.)

We may represent this branching structure with a matrix $Q=[q_{ij}]$ such that $q_{ij}=1$ if the $j$th event is a child of the $i$th event, and $0$ otherwise (note $q_{ii}=1$ only if $i$ is a parent event).  This provides a natural latent variable for the expectation-maximization procedure we outline in \textit{Methods}, and it also provides a source of information about the structure of an individual's day-to-day activity. 

The multivariate periodic Hawkes process (MPHP) provides a powerful framework to model the activity of an individual, capturing both interdependence of event types and periodicity.  The parameter $\delta$ may be fine tuned to represent different granularities of periodicity as the question requires or data permits.  For this study, we retain simplicity by assuming $\delta$ and $\omega$ are global parameters for an individual governing all activity types, but this could be fairly straightforwardly extended if appropriate.  Treating $\omega$ as a global parameter has precedent and significantly reduces the burden of parameter estimation.

\begin{figure}[tb]
    \centering
    \includegraphics[width=0.9\linewidth]{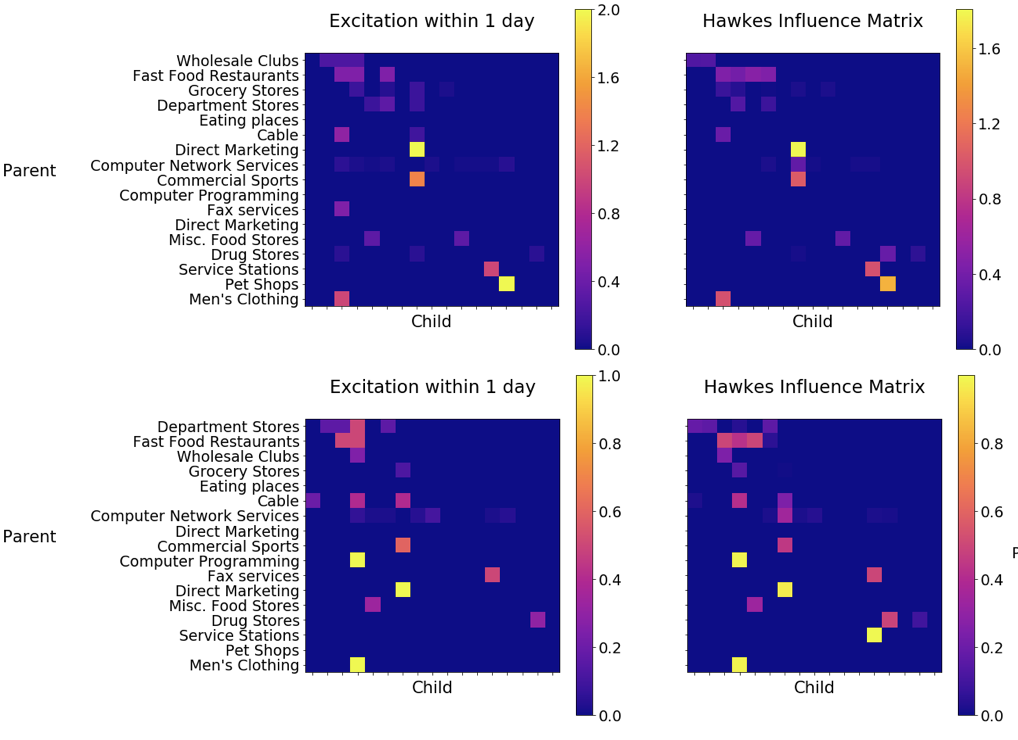}
    \caption{\textbf{Interdependence and co-excitation.}  For two sample users (\textit{top} and \textit{bottom}), we compare the same-day co-occurrence of pairs of transaction behavior in the empirical data (\textit{left}) with the excitation (or ``influence'') matrix $A$ estimated for the MPHP model (\textit{right}).  Note the close correspondence, with important exceptions: (i) where the Hawkes model identifies high co-occurrence behavior as spurious, assigning a low excitation parameter, and (ii) where despite low same-day co-occurrence, the Hawkes model detects important co-excitation.}
    \label{excitation}
\end{figure}

\begin{figure}[tb]
    \centering
    \includegraphics[trim={0cm 0cm 0cm 1cm}, clip, width=\linewidth]{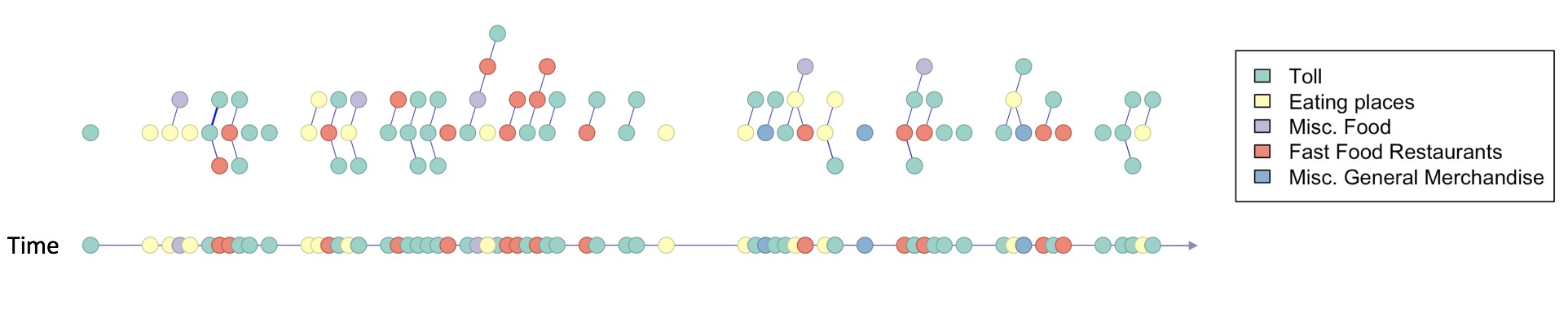}
    \caption{\textbf{Branching structure.} The EM parameter estimation technique uncovers a branching structure for an individual's behavior, encoded in the latent branching matrix $Q$.  A sequence of events for one individual is depicted using only time (\textit{bottom}) and depicting the uncovered branching structure (\textit{top}). We see, for this individual, frequent self-excitation among tolls, and a tendency for fast food purchases to act as ``parent'' events for other purchase activity.}
\label{branching}
\end{figure}

\begin{figure}[tb]
    \centering
    \includegraphics[trim={0 0.4cm 3.5cm 0cm}, clip, height=\linewidth, angle=270]{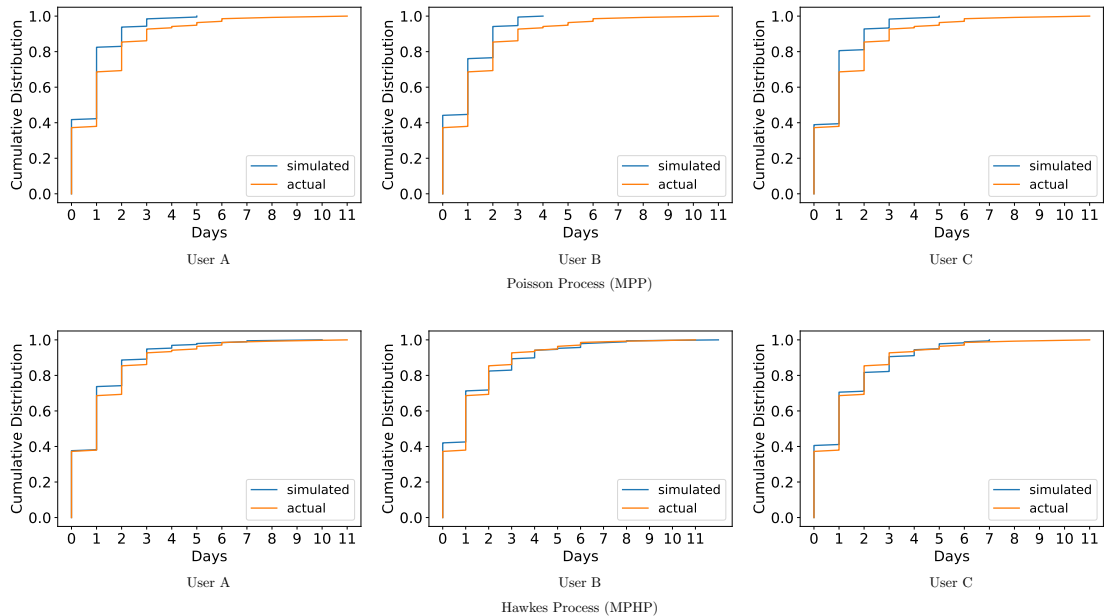}
    \caption{\textbf{Inter-event time distributions.}  We simulate event sequences using a multivariate Poisson process (MPP, \textit{top}) and a MPHP (\textit{bottom}) for three users, then compare the resulting CDFs of inter-event times with the empirical data. The empirical data exhibits heavy tails which the MPHP is able to capture but the MPP is not. At the 5\% significance level, MPHP can only be rejected for 29.8\% of individuals, MPP for 100\% of individuals.}
\label{inter-eventcdf}
\end{figure}

\begin{figure}[tb]
    \centering
    \includegraphics[width=\linewidth]{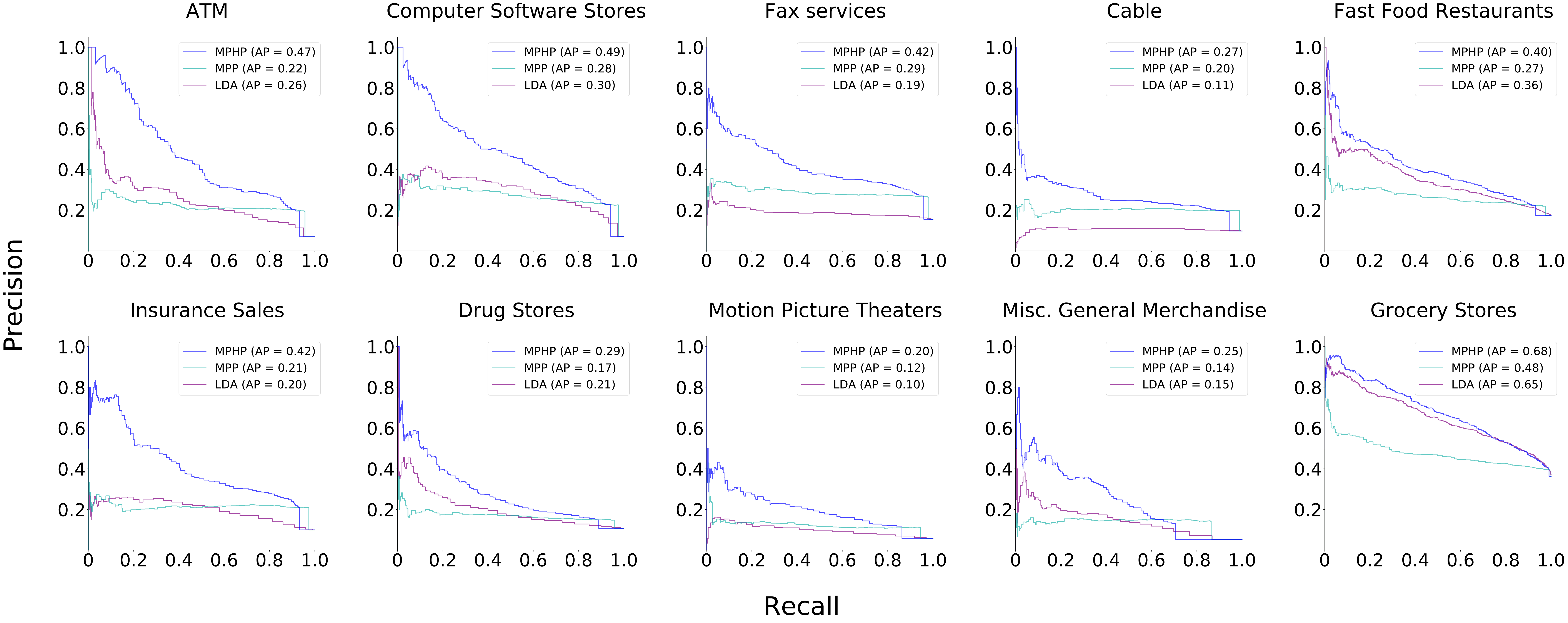}
    \caption{\textbf{Predictive ability.}  Precision-recall curves for predicting the occurrence of a particular activity in the next $\varepsilon = 2$ days are shown for the 10 most common activity types, for the MPHP (\textit{blue}), MPP (\textit{cyan}), and LDA (\textit{magenta}) models.  Average precision among all thresholds indicated in legend.  The MPHP outperforms both models in all activity types.}
\label{pred}
\end{figure}


\section{Results}
\label{sec:results}

\subsection{Data}

We apply our model to six months of credit card records in Mexico City.  In this dataset, the activity types are purchase categories (e.g. tolls, fast food restaurants, drug stores).  Our results focus on the 23,317 individual users within the dataset with at least 90 transactions.  The granularity of time stamps is one day, and thus we model periodicity depending on the day of week only.  Examining this data in Fig. \ref{userdata}, we see that empirical patterns in credit card transaction history show clear burstiness for two sample users.


\subsection{Interdependence and Branching Structure}

We begin with some qualitative observations about interdependence and branching structure of individuals' activity patterns encoded in the excitation matrix $A$ and branching matrix $Q$  revealed by the MPHP parameter estimation approach outlined in \textit{Methods}.  

Recall each excitation parameter $\alpha_{ij}$ in $A$ scales the additive contribution of previous event $i$ on the rate of arrival for event $j$, and thus describes the dependence of activity $j$ on activity $i$ --- for this reason, one may refer to $A$ as the ``influence matrix'' although we are careful to note we are making no claims of causality.  Fig. \ref{excitation} compares, for two sample individuals, the same-day co-occurrence of pairs of transaction behavior in the empirical data (\textit{left}) with the estimated excitation matrix $A$ of the MPHP model (\textit{right}).  We note a close correspondence between quantities, indicating $A$ primarily captures simple co-occurrence relationships.  We also, however, note important exceptions: (i) where the Hawkes model identifies high co-occurrence behavior as spurious, assigning a low excitation parameter, and (ii) where despite low same-day co-occurrence, the Hawkes model detects important co-excitation.

Also recall the latent variable $q_{ij}$ represents whether event $i$ was a parent of a child event $j$, and our EM methodology provides an estimate $p_{ij}$ such that $P=[p_{ij}]=\mathbb{E}[Q]$, as described in \textit{Methods}.  This reveals additional, branching structure in a sequence of transactions, depicted in Fig. \ref{branching}.  While the credit card records provide only a ``flat'' timeline of events, the MPHP learns tendencies of certain event types to act as ``parent events'' of others.  For example, for this individual, we observe frequent self-excitation among tolls, and a tendency for fast food purchases to act as parent events for other purchase activity.

\subsection{Monte Carlo Hypothesis Testing}

We next test the MPHP's ability to capture an individual's distribution of interevent times, in comparison with a non-homogeneous (periodic) Poissonian baseline.   For each user, we learn a MPHP model and compare the model's predictions with the empirical cumulative distribution of inter-event times (see  \textit{Methods}). Due to the inherent burstiness of human activity, we expect heavy tails in each distribution.  The proposed MPHP better captures bursty inter-event time distributions than a multivariate periodic Poisson process (MPP), as shown by the empirical distributions for several example shoppers in Fig. \ref{inter-eventcdf}.

Because the estimated parameters depend on the empirical data, we use Monte Carlo hypothesis testing to assess the significance of the agreement for each user. At the 5\% significance level, our model can only be rejected for 29.8\% of users. For this minority, the probability of a one day inter-event time was comparable to that of a same day inter-event time, indicating that their excitation function is not exponential decaying.  For these cases, a better fit may be achieved with the substitution of a triggering kernel that does not start decaying until after one day. A nonparametric triggering kernel \cite{nonparametrickernel,nonparametrickernel2} could result in a closer fit for all users, but would result in losses in both interpretability and scalability.

In comparison to the Hawkes model, a Poisson model, even with periodic intensity, is rejected for 100\% of users.  We see that the proposed model is complex enough to capture a wide range of human behavior, while remaining simple enough to lend insight into patterns within individual activity.

\subsection{Prediction}

Lastly, we compare the predictive ability of the Hawkes model with a periodic Poisson process and latent Dirichlet allocation (LDA) \cite{lda}.  LDA is a generative statistical model commonly used in the context of natural language processing.  It is able to identify shared patterns across users, without the temporal dimension.

To evaluate the predictive ability of our model, we consider a binary classification task: given all purchases of a user until time $t$, will the user make a new purchase type $i$ in the next time period, $[t, t+ \varepsilon]$? For each user, $t$ is a randomly chosen day within the last 10\% of the user's total history. We choose a small time window of $\varepsilon = 2$ days to measure each model's ability to capture self-exciting behavior in addition to general patterns in activity. 

For almost all activity types, an overwhelming majority of users will not make a purchase of that type (90\% - 97\%). Due to the imbalanced nature of the data, we use precision and recall as metrics to evaluate prediction performance.  Denote $P$ as the number of positives (where a positive indicates that the corresponding individual made a purchase of the specified activity type within the time window), $TP$ as the number of true positives (individuals correctly classified as making a purchase of the specified activity type), and $CP$ as those the algorithm classifies positive, correctly or incorrectly. Then we have $\text{Precision} = TP/CP$ and $\text{Recall} = TP/P$. 


For each model, we first estimate parameters for an individual using the time interval $[0,t)$, then repeatedly generate sequences from the model in the time window $[t, t+\varepsilon]$.  We record the percent of sequences containing a purchase of the specified type and compare with the data.  For a discussion of our sampling procedure for the MPHP, see \textit{Methods}.  MPP and LDA are standard techniques and we refer the interested reader to more detailed texts.   


As we can see from Fig. \ref{pred}, the multivariate periodic Hawkes process (MPHP) outperforms both the multivariate periodic Poisson process (MPP) and LDA in predicting a range of event types.  This shows that the temporal dimension is necessary to describe human behavior, and further, that more than periodicity is needed.


\section{Discussion}
\label{sec:discussion}

Our results demonstrate that the excitation structure between events, when coupled with weekly cycles, is able to generate realistic and predictive trajectories of shopping activity. In addition to accurately describing heavy tails in inter-event time distributions, the proposed model solidly outperforms baseline models in difficult prediction tasks. In general, our results demonstrate the effectiveness of task prioritization and periodicity in explaining activity involving credit card transactions, however, the model is readily applicable to a broader range of individual human activity (for example, making phone calls, doing chores) which we hope to pursue in future work.


Furthermore, the model is highly interpretable, in contrast with many state-of-the-art generative and predictive models.  The excitation matrix $A$ gives a direct encoding of the interdependence of activity types.  The branching matrix $Q$ revealed through our proposed EM technique provides more than a convenient latent variable to aid in parameter estimation, it gives important information about the relationships between specific events in an individual's activity history.  This provides an interesting direction for future research: in past work \cite{rdicle}, communities have been found that display consistent behavioral trends in terms of spending and demographics. We can further characterize the behavior of these communities by including the temporal dimension, thus describing temporal lifestyles at urban scale.


Lastly, due to its generality and flexibility, the Hawkes model is well-suited to describe a wide range of phenomena, with applications in diverse fields as noted in the \textit{Introduction}.  The addition of periodic effects, and the tractability of our proposed MAP EM parameter estimation technique, provide valuable extensions in this already expansive realm of applications.



\section{Methods}
\label{sec:methods}

In this section we present our methodologies for parameter estimation, simulation, and hypothesis testing of the multivariate periodic Hawkes process (MPHP)


\subsection{MAP Expectation-Maximization} 

Consider again a sequence of $N$ events $\tau=\{(t_i, u_i)\}_{i=1}^N$ over a time period $[0,T]$, where $t_i$ is the time of the $i$th event and $u_i$ is the event type, out of $U$ possible types.  Also, recall the conditional intensity $\lambda(t)$ of the MPHP as described in Eq. \ref{eq:mphp}, with parameters $\Theta=(\mu, \delta, A)$. 

The form of $\lambda(t)$ actually allows us to work out the likelihood function $p(\Theta | \tau)$ in closed form.  This fact permits direct parameter estimation of $\Theta$ via maximum likelihood estimation methodologies.  However, in practice, such methods pose many challenges due to the objective function's low curvature and large parameter space, requiring the invocation of strong regularization schemes and sophisticated optimization techniques \cite{zhou2013learning}.  Alternatively, the additive nature of the intensity permits hierarchical Bayesian approaches with priors achieving the necessary regularization, but which then require variational or sampling-based approaches \cite{linderman2017bayesian}.

We introduce a simpler approach that still incorporates some regularization in the form of a prior on $A$ and $\delta$.  We propose an extension of the expectation-maximization (EM) scheme presented in \cite{zipkin2016point,Veen2008} to the multivariate periodic case, and we incorporate a prior on the parameters governing interaction and periodicity through maximum aposteriori (MAP) EM.\footnote{See \url{https://github.com/stmorse/hawkes} (MHP) and \url{https://github.com/sharonxu/hawkes} (MPHP) for implementations of this methodology.}  

First, recall the interpretation of the Hawkes model as a branching process as in Fig. \ref{branching}.  We introduce latent variables $Q=[q_{ij}]$ over the data $\tau$, called the \emph{branching matrix}, defined such that $q_{ij}=1$ if the event at $t_i$ was caused by the event at $t_j$ (0 otherwise), and note $q_{ii} = 1$ implies the event at $t_i$ was a background event.  We will estimate these latent variables by computing their expected value $P=\mathbb{E}[Q]=[p_{ij}]$. 

We may now define the (expected) complete data log-likelihood as
\begin{align}
\mathbb{E}\big[\log p(\tau, Q | \Theta) \ | \ Q\big] = 
&\sum_{i=1}^N p_{ii} \log \frac{\mu_{u_i}}{p_{ii}} + 
\sum_{i=1}^N \sum_{j=1}^{i-1} p_{ij} \log \frac{\alpha_{u_i u_j} g(t_i - t_j)}{p_{ij}} \nonumber \\
&- \dfrac{T}{D} \sum_{u, d} \mu_u \delta_d
- \sum_{u=1}^U \sum_{j=1}^N \alpha_{uu_j} G(T-t_j)
\label{eq:completeloglikelihood}
\end{align}
where $G(t)=\int_0^t g(s) ds$.  See \cite{Veen2008,zipkin2016point,zhou2013learning} for a more thorough treatment of the derivation of Eq. \eqref{eq:completeloglikelihood}.

The complete data log-posterior is
\begin{equation}
\log p(\Theta | \tau, Q) \propto \log p(\tau, Q | \Theta) + \log p(\Theta)
\label{eq:logposterior}
\end{equation}
We seek to maximize this posterior, subject to the periodicity constraints $\sum_d \delta_d = D$ and $\sum_h \rho_h = H$, using the expectation-maximization (EM) algorithm.

By using a MAP estimate, we have the opportunity to place a prior on the excitation matrix entries $A=[\alpha_{ij}]$ and the periodicity scaling parameters $\delta$.  A Gamma prior is conjugate with the Poisson distributions of the complete data likelihood,
\begin{align}
\alpha_{i,j} &\sim \text{Gamma}(\alpha_{ij}; \ s_{ij}, t_{ij})\\
\delta_d &\sim \text{Gamma}(\delta_d; \ w_{d}, x_{d})
\end{align}
Using Gamma priors also provides a nice interpretation of the hyperparameters as ``pseudocounts'' --- for example, in the case of $\alpha_{ij}$, they represent already observed counts of parent and child events between the pair $(ij)$.

The EM algorithm alternates between finding the expected value of $P=[p_{ij}]$ of $Q$ in the expectation step (E-step), and maximizing the posterior with respect to $\Theta$ in the maximization step (M-step).  More formally, the E-step computes $P^{(k+1)} = \mathbb{E}[Q|\tau, \Theta^k]$, 
\begin{align}
p_{ii}^{(k+1)} &= \frac{\mu_{u_i}^{(k)}\delta_{d_i}^{(k)}}
{\mu_i^{(k)}\delta_{d_i}^{(k)}
+ \sum_{j=1}^{i-1}\alpha_{u_i u_j}^{(k)} g(t_i-t_j)}\\
p_{ij}^{(k+1)} &= \frac{\alpha_{u_i u_j}^{(k)} g(t_i-t_j)}{\mu_i^{(k)}\delta_{d_i}^{(k)} + 
\sum_{j=1}^{i-1}\alpha_{u_i u_j}^{(k)} g(t_i-t_j)}   
\end{align}
where both formulas follow directly from the additive property of Poisson processes.

For the M-step, we may obtain update formulas by explicitly solving the stationarity condition $\partial/\partial\Theta = 0$ on \ref{eq:logposterior}:
\begin{align}
\mu_u^{(k+1)} &= \frac{\sum_{i: u_i=u} p_{ii}^{(k)}}{T}\\
\alpha_{uu'}^{(k+1)} &= \frac{\sum_{i: u_i=u} \sum_{j: u_j=u', j<i} p_{ij}^{(k)} + s_{uu'} - 1}{\sum_{i=1}^N \sum_{j: \ u_j=u', j<i} G(T-t_j) + t_{uu'}}  \\
\delta_d^{(k+1)} &= \frac{\sum_{i: d_i=d} p_{ii}^{(k)} + w_d - 1}{\dfrac{1}{D} \sum_i p_{ii} + w_d - 1}
\end{align}
These formulas also have accessible interpretations; for example, the estimate for the background rate $\mu_u$ is the number of background events in $u$ divided by total time $T$. Also note the role of the Gamma hyperparameters as pseudocounts for child and parent events.  Finally, note we may take advantage of the fact that $G(T-t_j)\approx 1$ to simplify calculation of $\alpha_{uu'}$.

\subsection{Simulation}

In order to sample event sequences from the process, we use the thinning method due to Ogata \cite{ogata1981}.  We make two important modifications to this algorithm that increase efficiency; our method is described fully in \ref{algo:mphp}.

As typically described \cite{Toke2011,Laub2015}, the thinning method requires $O(n^2 U^2)$ operations to draw $n$ samples over $U$ dimensions, which is prohibitive for large processes. Instead, we modify an approach mentioned in \cite{Valera2015}.  Namely, given the rates at the last event $t_k$ we can calculate $\lambda(t)$ for $t>t_k$ by
\begin{equation}
\lambda_u(t) = \mu_u\delta_d + e^{-\omega(t-t_k)}\big(a_{uu'_k} \omega + (\lambda_u(t_k) - \mu_u\delta_d)\big)
\end{equation}
which we can do in $O(1)$, and only requires saving the rates at the most recent event. Secondly, we improve the typical attribution/rejection framework \cite{Toke2011,Laub2015} for each activity type by instead viewing the procedure as a weighted random sample over activity types, that is, the integers $1, \cdots, U$.  This allows us to forgo for-loops and instead use optimized functions for weighted random samples.

\subsection{Monte Carlo Hypothesis Testing}

Following \cite{poissonianemail}, we assess statistical goodness of fit of a model using the area statistic, or the area between two cumulative distribution functions. We use the area statistic to compare the interevent-time distributions of the empirical data and event sequences simulated from the model. 

We estimate the parameters $\Theta_1$ of a stochastic process from some dataset $\mathcal{D}$, use these to simulate data $\mathcal{D}_1$, and record the area $A$ between interevent CDFs of the real and simulated datasets.  We then estimate parameters $\Theta_2$ from $\mathcal{D}_1$, simulate data $\mathcal{D}_2$, and record the area $A_0$ between interevent CDFs of $\mathcal{D}_1$ and $\mathcal{D}_2$.  We repeat this $M$ times and compute the test statistic between the groups $\{A_m\}$ and $\{A_0^m\}$.

{\LinesNumberedHidden
\begin{algorithm}[tb]
\DontPrintSemicolon
  \vspace{0.5em}
  \KwIn{$\mu = \{ \mu_v \}, \; A = (a_{ij}), \; \omega_1, \; \omega_2, \; \delta = \{ \delta_d \}$, horizon}
  \KwOut{Sequence of event types $\{(t_i, v_i)\}_{i=1}^N$}
  Simulate first event:\;
  $I^* \leftarrow \sum_v  \mu_v$ \;
  $D^* \leftarrow \sum_{i=1}^7(\delta_d)$\;
  \Repeat{
$U < \dfrac{d_0}{\sum_{d=1}{7} \delta_d}$
  }{
  $t_0 \sim \mathrm{Exp}(1/M)$\;
  $d_0 \leftarrow \mathrm{day of week} (t_0)$\;
  $U \sim \mathrm{Unif}(0, 1)$
  }  
 
    $v_0 \leftarrow v $ w.p. $ \mu_v/I^*$ \;
     $\lambda(t_0) \leftarrow \mu \delta_d $ \par
  
  \textbf{General procedure:}
$k\leftarrow 0$\\
  \textbf{Step 1}\\
    \quad $I^* \leftarrow \mathrm{max}(\delta_d)\sum_{\mu_v}\mathrm{max}(\delta_d) \lambda_v(t_k) + \omega \sum_{v}a_{vv_k}$\\
  \textbf{Step 2}\\
  \quad $t' \leftarrow t_k + s, \ s\sim \text{Exp}(1/I^*)$\\
    \If{$t' > \mathrm{horizon}$ }{
         return $\{(t_i, v_i)\}$
    }
  $\lambda(t') \leftarrow \delta_{d'} \mu + e^{-\omega(t'-t_k)}\big(A_{v_k}\omega + \lambda(t_k) - \delta_{d_k}\mu\big)$\\
  \textbf{Step 3}\\
    $u' \leftarrow u$ w.p. $\mu_v/ I^*$\\
    $d' \leftarrow d_k$\\
    \uIf{$v'$ is $v+1$}{Reject}
    \uElse{
    	Attribute:\\
        \quad $t_{k+1} \leftarrow t'$\\
        \quad $v_{k+1} \leftarrow v'$\\ 
        \quad $d_{k+1} \leftarrow d'$ \\
        \quad $\lambda(t_{k+1}) \leftarrow \lambda(t')$\\
        \quad $k \leftarrow k+1$\\
        Step 1
}

\caption{Simulation of event sequences from a MPHP}\label{algo:mphp}
\end{algorithm}
}



\clearpage

\bibliographystyle{plain}  
\bibliography{biblio}

\end{document}